# Analysis of Digital Sovereignty and Identity: From Digitization to Digitalization

Kheng Leong, Tan, Chi-Hung, Chi, and Kwok-Yan, Lam, *Senior Member, IEEE*

***Abstract*— Advances in emerging technologies have accelerated digital transformation with the pervasive digitalization of the economy and society, driving innovations such as smart cities, industry 4.0 and FinTech. Unlike digitization, digitalization is a transformation to improve processes by leveraging digital technologies and digitized data. The cyberspace has evolved from a hardware internetworking infrastructure to the notion of a virtual environment, transforming how people, business and government interact and operate. Through this transformation, lots of personal data are captured which individuals have no ownership or control over, threatening their privacy. It is therefore necessary for the data owners to have control over the ownership, custody and utilization of their data and to protect one’s digital assets and identity through proper data governance, cybersecurity control and privacy protection. This results in the notions of data sovereignty and digital sovereignty - two conceptually related terms, but different focuses. This paper first explains these two concepts in terms of their guiding principles, laws and regulations requirements, and analyse and discuss the technical challenges of implementing these requirements. Next, to understand the emerging trend shift in digital sovereignty towards individuals to take complete control of the security and privacy of their own digital assets, this paper conducts a systematic study and analysis of Self-Sovereign Identity, and discuss existing solutions and point out that an efficient key management system, scalability and interoperability of the solutions and well established standards are some of its challenges and open problems to wide deployments.**

## I. INTRODUCTION

Advances in emerging technologies, including 5G telecommunication, Internet of Things (IoT), data analytics and artificial intelligence (AI) have resulted in the acceleration of digital transformation with the pervasive adoption of digitalization of the economy and society. The cyberspace has evolved from a hardware internetworking infrastructure to the notion of a virtual environment where people, business and government interact by digital means. Through the decades, cyberspace has become an enabler of technology innovations such as smart cities, industry 4.0 and FinTech; it rapidly transforms the way people, business and government interact and operate, and accelerates the digitization to digitalization movement in the process. Digitization refers to the process of encoding pieces of data into digital formats, turning analogue data into computer readable format, for transmission, re-use and information processing. This digitization process has increased exponentially the amount of data that could be collected and processed. Very often these data are our personal data closely related to our online activities and profile behaviors which are mostly identifiable via our online identity. From a more general point of view, the act of digitization does not by itself involve digitalization in the initial novel intention of digital transformation. Digitalization is an eventual technological trend that leverages digital technologies to transform the data into a quantified format, turning many aspects of our life into data which is subsequently transferred into information realised as a new form of value. As the cyberspace interwoven into the fabrics of people’s daily lives, a lot of personal data are captured, which individuals are unable to take ownership or control over. These data are subsequently being made use of by third party data collectors, through data analytics and AI for value creation. But what is the problem? The problem is that there is no consensus from the data owner, the individual which the data is about, on how data should be and can be used! Data owner loses his/her control of the data. The use of the data might not be correct, or might not be in favor of the owner’s initial intention or context, for example, fake news that is being manipulated or taken out of context. In other words, the data owner does not have “content right” to decide the destiny of his/her data. By content right, this paper refers to the fact that the data owner might not agree with the way data is used, interpreted and shared. This digitalization is mostly initiated by big tech companies, which control the online platforms and data registries, like social and cloud platforms, with user data readily available to them for the purposes of data analytics and artificial intelligence adaptations for their business operations. These data are individual personal information in the form of personally identifiable information (PII), online profiles, activities and behaviors. It is an individual’s digital shadow relating to his/her life in cyberspace in which s/he would like to preserve as his/her own privacy and self-determine its access –

Kheng Leong, Tan is with the School of Computer Science and Engineering, Nanyang Technological University, 50 Nanyang Avenue, Singapore 639798 (e-mail: tank0260@ntu.edu.sg).

Chi-Hung, Chi is with the Strategic Centre for Research in Privacy-Preserving Technologies, Nanyang Technological University, 50 Nanyang Avenue, Research Techno Plaza (RTP) BorderX Block, Level 4, Singapore 637533 (e-mail: chihung.chi@ntu.edu.sg).

Kwok-Yan, Lam is with the School of Computer Science and Engineering and the Strategic Centre for Research in Privacy-Preserving Technologies, Nanyang Technological University, 50 Nanyang Avenue, Research Techno Plaza (RTP) BorderX Block, Level 4, Singapore 637533 (e-mail: kwokyan.lam@ntu.edu.sg).

to have sovereignty over the data.

This problem of losing control is not easy to be solved as it involves people's change of mindset from securing data to opening up data for sharing and collaboration purposes. Even worse, the data that people open up more and more are very sensitive personal data. And the relevancy and traceability of the analyzed result to an individual data owner become more and more of concerns. Thus, under this digitalization movement, there are at least the observations below about the involved data that this work can make:

• People's expectations, as an individual, on the control and usage towards their own data, from keeping their data fully confidential to actively sharing their own data (example in social network).

• Collaboration and willingness of people in sharing of their data for analysis or machine learning usage for the discovery of insight and better prediction. For example, public data contributed by governments like data.gov.xx.

• People ultimately want to have ownership and control over the use of their own data, the autonomy in their roles as employees, consumers, and users of digital technologies and services. And the word "use" has a wide scope of meaning, from pure data access to the interpretation of data by third parties.

From computer science research perspectives, a shift of focuses is observed, from security to privacy to sovereignty, to match the change of people's mindset.

**Security**: This often refers to as data security and in essence, means protecting digital data, such as those in a database, from destructive forces and the unwanted actions of unauthorized users, such as a cyberattack or a data breach. Its goal focuses mainly on data access from unauthorized access, corruption, or theft throughout its entire lifecycle.

**Privacy**: Privacy is the claim of individuals, groups, or institutions to determine for themselves when, how, and to what extent information about them is communicated to others [1]. Individuals or groups can protect and preserve their privacy by secluding themselves or withholding information about themselves and thereby express themselves selectively.

**Sovereignty**: Digital sovereignty is commonly used by governments to convey the idea that states should reassert their authority over the internet and protect their citizens and businesses from the manifold challenges posed by digital transformation and the global technical infrastructure of the internet (especially the emergence of cloud). Governments have made it possible to enforce national laws and undertake governmental interventions in the digital sphere to address the issues of data and digital governance, and exercise sovereignty over the digital data. On the technological front, Google Cloud has started an initiative to provide the highest levels of digital sovereignty, conforming to Europe's terms, to enable Europe's businesses and organizations to exercise autonomy over their data [2].

With the widespread of digital literacy in smart nations, digital sovereignty has become an increasing demand from people's ownership, custody, and control over their own digital assets to the inalienable user data rights of exercising self-determination on their data destiny. It is observed that such shift of empowering an individual to have complete control over their digital data, including his/her discretion on what to share and with whom, is often built on top of the concept of digital identity [3]. A digital identity is a set of validated digital attributes and credentials for the digital world, similar to a person's identity for the real world [4-7]. It forms the basis for authentication and authority in cybersecurity [8-13]. The demanding sovereignty on digital identity results in the research on "Self-Sovereign Identity" (SSI), and this is currently gaining much research attentions and interests.

This paper will focus on the sovereignty aspects of the digitalization transformation and perform a comprehensive systematic literature review and study on data, digital sovereignty and self-sovereign identity. This paper's contributions are summarized as follows:

1) Aims to properly define the notion of digital sovereignty and provide an understanding of the nature of digital sovereignty problem and its importance in supporting digital transformation of the society/economy. This paper points out the ongoing emphasis and existing research, which involve from policymaking, regulation and enforcement, to the technical challenges of implementing the requirements in the notion of sovereignty.
2) Introduces data sovereignty concept which is often related to indigenous governance on the right of indigenous peoples. This provides a more complete and holistic view of sovereignty from the early days of the wishes and desires of the indigenous people to have data sovereignty to the challenges in execution of the sovereignty rights.
3) Proposes a systematic method of searching and selecting publications, which is guided by research questions. Finally, research questions and open challenges are discussed, which are useful for researchers and practitioners working on digital sovereignty and particularly SSI.

The remainder of the paper is organised as follows: Section II and III present a brief systematic review on data sovereignty and digital sovereignty respectively. Section IV presents a detailed systematic review of SSI covering the research methodology adopted, the research questions and their corresponding results, and finally an analysis and discussion on the results. Section V discusses the related work and Section VI concludes the paper.

## II. DATA SOVEREIGNTY

The notion of data sovereignty is actually not new but commonly focused on the struggle of indigenous peoples to reclaim sovereignty over their land, culture and heritage. Data sovereignty, according to Wikipedia [28], "is the idea that data are subject to the laws and governance structures within the nation it is collected". Earlier research was mobilized by indigenous scholars and from an indigenous perspective to voice the rights of indigenous peoples concerning the collection, ownership and application of data about their people, lifeways and territories" [18]. A survey by [19] shows the trend of the researches on the topic of "data sovereignty" and the growing interests in this area before and since 2015 till 2018. In the stated period, the academic and non-academic literature contributions are 89 and 2459 respectively. This paper also did a search in Proquest central on the keyword "data sovereignty" from 2018 till 2021 and the result shows a collection of 263 academic peer-reviewed, full-text papers. The larger increase shows the trend in this research area. A word cloud was generated on a sample of the 263 papers and the result is shown in Fig. 1. Prominent words observed from the word cloud are 'indigenous', 'control', 'access', 'policy', 'personal' and 'protection' which aligned with initial researches which situated on the larger struggle and wishes of indigenous peoples, especially in the Anglo settler–colonial CANZUS countries (Canada, Australia, New Zealand and the United States) [9], to reclaim sovereignty over their land, body, and culture. The movement toward open data and open science does not fully engage well with the indigenous people's rights and interests. Their voices and rights were missing concerning the "collection, ownership and application of data about their people, lifeways and territories" [21]. There was a need to establish indigenous frameworks and executable principles on the collection and/or creation of this data, assert greater control over the application and use of indigenous data and knowledge for collective benefit, and facilitate increased data sharing among entities. This provides the initial humanity and sociological science research perspectives on data sovereignty.

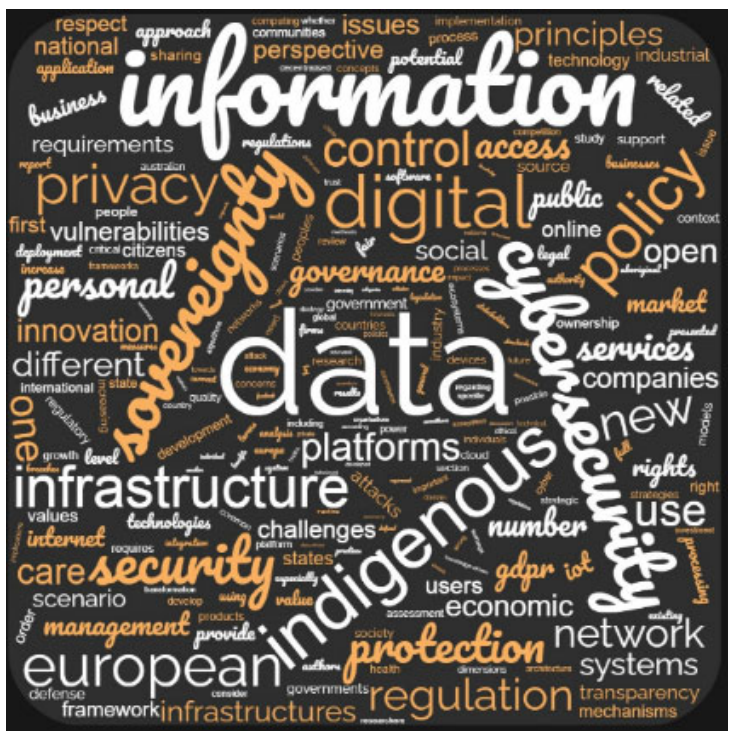

**Fig. 1.** Word cloud of "Data Sovereignty" papers

With the advance in technologies and access to the internet, coupled with the digital transformation movement of digitization to turn analogue data to digital, the discussion on data sovereignty is expanded to cover a wider area. According to the World Bank [22], by 2022, global internet traffic is expected to reach 150,000 GB of traffic per second, a 1,000-fold increase compared with the 156 GB in 2002. As organizations become increasingly global and interconnected, issues surrounding data security and privacy have become more complex, questions like where does that data really reside, who owns it, and what regulations should it be subject to, become critical considerations for organisations in compliance with the regulations of every market in which business is conducted. In short, data will be governed by regulations specific to the region in which it originates. Failing to meet the compliance can result in hefty fines or worse. This was evident in the word cloud from words like 'governance', 'privacy', 'regulation', 'security', 'platforms', 'infrastructure' and 'cybersecurity'.

So what is data sovereignty? There are two aspects to data sovereignty. The first aspect is on principles. A discussion on data sovereignty will inevitably relate to the discussions on indigenous people's rights and wishes to have control over the access of their data and two principles: the CARE and FAIR principles. The CARE Principles for Indigenous Data Governance were created to advance the legal principles underlying collective and individual data rights in the context of the United Nations Declaration on the Rights of Indigenous Peoples (UNDRIP) [23]. CARE is an acronym that stands for Collective Benefit, Authority to Control, Responsibility, Ethics [24]. It is the first attempt to outline collective rights as part of openness. It provides external data stakeholders with guidance and advice on governance practices and stewardship responsibilities for indigenous data. While CARE can be considered as part of the open data movement, its principles are still abstract ideals and there are still needs to develop criteria and tools to implement it and build on with other standards such as FAIR (findable, accessible, interoperable, reusable) [25] by considering power differentials and historical contexts. Developed in the Netherlands in 2015, the FAIR principles have since been taken up across the Western world as a way of sharing data that will maximise use and re-use. The rationale is that making data FAIR will support data and knowledge integration and promote sharing and re-use of data. Compare to CARE, the FAIR principles are more aligned to technical principles from an implementation perspective, relating how data can be searched, retrieved, shared, transferred and reused across different technological IT systems, platforms and environments. Technology will be a critical tool and infrastructure to operationalize these principles to exercise indigenous people's sovereignty and self-determination. Thus, the principles for indigenous data governance require enacting FAIR, but with CARE.

The second aspect is on law. Data Sovereignty is closely linked to the national law, regulations and perspectives of the countries where the data reside. It refers to the concept that the data an organization collects, stores, and processes are subject to the nation's laws and general best practices where it is physically located. Another closely related term is data residency. Data residency is when a business or government specifies the geographical location where its data should be stored. Thus, at its core, data sovereignty is about protecting sensitive, private data and ensuring it remains under the control of its owner within the specified country. In layman's terms, this means that a business has to store the personal information of its customers in a way that complies with all the data privacy regulations, best practices, and guidelines of the host country. With regulations like the European Union's General Data Protection Regulation (GDPR) [26] setting the bar for data privacy protection, it's more important than ever for organisations to proactively safeguard their sensitive customer and employee data, everywhere it's stored and shared. In the U.S., the California Consumer Privacy Act (CCPA) [27] has a similar objective to give California residents greater control over how their data is used and stored.

The principle of the law on data sovereignty is that data belongs to the jurisdiction of the nation-state where it is originally held in binary form. Earlier researches on data sovereignty are mainly focusing on the humanity and sociological aspects conducted by researchers in that line of fields who work with the data (or asset) owners and defined suitable data sovereignty principles like CARE and FAIR. However, the principles lack implementation and technical details to execute them as the researchers involved in defining them may lack the computer science background to architect a technical solution. The same goes for the defined or legislated laws revolving around data sovereignty and intended protection, which will also require technology supports to enforce them and execute the regulations and policies. Therefore, with the lack of operational details, implementation mechanisms and enacting systems to execute and realize the data sovereignty principles and law, it is a challenge to govern the stewardship and application of data to fully assert sovereignty on the locally hosted data of the people which the laws are designed to protect. With digital transformation that enables data to transcend geopolitics and economics, nation-states will need to reassert their authority to protect their citizens and businesses and look into data sovereignty on a wider digitalized aspect, that is, digital sovereignty. Thus, an understanding of what is digital sovereignty and the existing research in this area is necessary.

## III. DIGITAL SOVEREIGNTY

Digital sovereignty is a key idea in this digitalization age. Digitalization has become a major driver that transcends geopolitics and economics with waves of technological innovations, especially in artificial intelligence. This acceleration in digitalization or digital transformation is in realization of the power of data in which a complex ecosystem of entities has emerged to collect, analyse and trade the value that may be extracted from it. Through this transformation, lots of personal data are captured which individuals have no ownership or control over, threatening their privacy. It is therefore necessary to have control over the ownership, custody and utilization of these data and protect one's digital assets and identity through proper data governance, cybersecurity control and privacy protection. Thus, the notion of digital sovereignty.

According to the definition drafted by the EU Federal Chancellery [22]: "Digital sovereignty describes the ability to shape the digital transformation in a self-determined manner with regard to hardware, software, services, and skills. Being digitally sovereign does not mean resorting to protectionist measures or doing everything yourself. Being digitally sovereign means, within the framework of applicable law, making sovereign decisions about the areas in which independence is desired or necessary."

A similar survey by [19] shows the research on the topic of "digital sovereignty" and the trend in the interests of this topic. Before and since 2015 till 2018, the academic and non-academic literature contributions are 22 and 239 respectively. This paper did a similar search in Proquest central on the keyword "digital sovereignty" from 2018 till 2021 and it resulted in a collection of 39 academic peer-reviewed, full-text papers. This is a more proportional increase as compared to the data sovereignty figure presented earlier. This can be due to the level of understanding of digital sovereignty versa data sovereignty among non-academics and their coverage. Compared to data sovereignty, digital sovereignty is still a relatively new topic area. A word cloud on a sample of these 39 papers is shown in Fig. 2. A glance at the word cloud can reveal governments, national, states related words like 'commission', 'Europe' ('European'), 'political' and 'regulation' which align with a nation's desires to have sovereignty over the control and autonomy of the data. Other focuses are prominent words like 'technology', 'security', 'internet', 'infrastructure', 'services', 'cloud', 'platforms', '’privacy' and 'cybersecurity'. It can be observed that the discussions and focuses of the recent papers have shifted into the execution of digital sovereignty relating to technologies, infrastructures and cloud platforms. There are also security and cybersecurity concerns relating to privacy and data integrity as a result of artificial intelligence applications [28], aligning towards computer science research. In the gathered papers, EU is commonly cited and linked to the efforts they have made on asserting their digital sovereignty with implementations of regulations. A commonly cited regulation, in supporting digital sovereignty requirement is the European GDPR which grants and protects privacy rights to individuals located in the EU when their personal data is processed by non-EU companies that offer goods or services to them or monitor their behaviors. Most of the gathered papers provide digital sovereignty from the government and regulatory perspectives with the government as the central authority to protect its citizen's privacy and data rights. On closer examination of the word cloud, word like 'individual' is observed, which may suggest the research also focuses on individuals' human rights to have personal control on the ownership and ethical use of their data. Co-relating to the word cloud generated for data sovereignty in Fig 1, the words 'personal', 'control' and 'ownership' are also visible, suggesting the need of individuals to have sovereignty of their data, whether in analogue or digitalized form. Thus, the concepts of digital sovereignty are not limited to the control of the state over the use and design of critical digital systems, and the data generated and stored therein, but also for its people, on the individual level [19], to regain custody and control over the security and privacy of their own data. From a computer science research point of view, digital sovereignty requirements need to be aligned with data security and privacy issues [29-31] and require adopting information security triad principles; confidentiality, integrity and availability. Digital sovereignty on an individual basis cannot be ignored. Therefore, digital sovereignty can be viewed from a cybersecurity perspective and focus on two

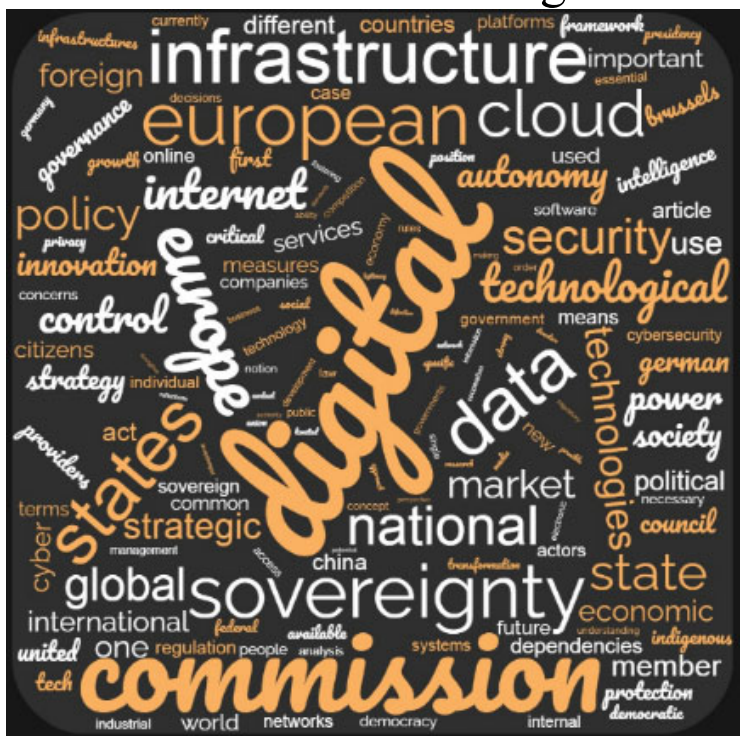

**Fig. 2.** Word cloud for "Digital Sovereignty" papers

aspects; at the national level as well as individual [27, 32-34].

### *A. National Level*

On a nation-state aspect level, the term digital sovereignty has been used by governments to convey the idea that states should reassert their authority over the internet and protect their citizens and businesses to control the digital infrastructure on their territories and the data of their people. With the dominant position of big tech companies also known as 'GAFAM' (Google, Apple, Facebook, Amazon, Microsoft) in the field of cloud computing and social media, data of citizens and companies are by now virtually stored and utilised in the cloud of these big tech companies. Nation-states of non-big tech origins would therefore like to reassert their authority [35] and control over the data and its usage. In the case of EU, in recognizing the importance of retaining this sovereignty and to constraint the 'platform power', the EU has channeled its efforts into tighter, more comprehensive regulation of the tech sector as a full-blown defense of Europe's digital sovereignty, becoming the world's leading "regulatory superpower" [33]. In addition, governments in the European Council, most specifically German, announced their intention "to establish digital sovereignty as a leitmotiv of European digital policy" [32]. The move attempts to reassert the government's authority over cyberspace [25] and protects their citizens and businesses from the challenges of not having control of their data. Two notable EU project initiatives are the GAIA-X cloud initiative initiated by Germany, with the support of France, and the European Cloud Federation initiative. The GAIA-X cloud initiative, launched in 2019 [36] whose origin stems from the German Federal Government, aims to create its own European offering of cloud infrastructure, services, and data is explicitly based on principles of sovereignty-by-design, where the customer has the full control over the storage and processing of the data and access thereto. The infrastructure aims to meet the highest standards in terms of digital sovereignty and foster innovation. GAIA-X is in line with the existing EU's GDPR which establishes key requirements for data handling related to European individuals or businesses. The European Cloud Federation initiative sets the standards for interoperability between providers and portability of data, where cloud providers will be expected to offer a choice as to where (personal) data are stored and processed, without otherwise requiring storage in Europe. From an initiation of digital sovereignty perspective, the GAIA-X project is a viable and promising initiative but requires a high degree of commitment and coordination from the EU member states which can be a challenge. With the lack of binding European cloud policies, currently the cloud choices made by the EU member states are through outsourcing which is only subject to the specific requirements of their own national cloud frameworks. Outsourcing also brings about the risk of vendor lock-in that threatens digital sovereignty [37]. The major market players offer limited interoperability and portability of data and applications. They can use their own standards and build their own private internet infrastructure, making any interconnection difficult, both in terms of infrastructure and data exchange. In addition, dependence on foreign providers brings with it control from other countries, which have different rules of play concerning espionage, privacy, and government access to data. From a cybersecurity perspective, several risks arise that can undermine digital sovereignty, such as the lack of control over the platforms and environment where the data reside and the level of data protection, which is often at the discretion of the providers based on the service level agreement (SLA) which the nation-states sign. Nation-states can inevitably face a spectrum of threats to their data from systematic theft of intellectual property, digital extortion, targeted misinformation, and systematic infiltration of social media [34] to influence elections and democratic processes. For digital sovereignty at the national level, in order to ensure the controls of the digital assets the nation-states seek to protect, there is a need for them to enhance their data infrastructure and the design of digital policy that can contribute to the development and execution of digital governance. At the same time, they also need to cope with its geo-economic and geopolitical implications to foster multilateral cooperation [35] at the international level.

### *B. Individual*

Digital sovereignty claims are not only the wishes of nations but is also the wishes of their citizen as individuals. Emphasizing the importance of individual self-determination, these claims focus on the autonomy of citizens, in their roles as employees, consumers, and individual users of digital technologies and services, to be able to determine who gets to access and use the data about them as individual [32]. An interesting aspect of this is the departure from a state-centered understanding of sovereignty. Instead of viewing sovereignty as the prerequisite to exercise authority in a specific territory controlled by government regulation and policies, digital sovereignty is also viewed as the ability of individuals to take actions and decisions in a conscious, deliberate and independent manner over the access and handling of his/her data - the self-sovereignty over one's data [27]. Thus, from the citizens' perspectives as individuals, digital sovereignty can be applied on an individual basis with the bottom line consideration of how their personal data and digital assets (PII) are treated and as an individual user of digital technologies and services [31, 32].

Digital transformation has brought data to the core of innovation that requires the sharing, exchanging and learning of the data that transcended geopolitics and economics, for example, in the area of modern artificial intelligence (AI) [28] or internet of things (IoT) [38]. The data being shared or utilized by these innovative services are often personal and sensitive in nature which individuals seek to control its level of access and disclosure of details. Individuals being the data owner, should have the sovereignty to control and authorize the usage and access of their data [33]. The digital transformation and utilization of these personal and sensitive data challenge the sovereignty of the individual and one's privacy. With clouds being the common platforms for the storage and access of data and mostly controlled by the 'GAFAM' big tech companies, there are significant concerns about the handling of these particularly sensitive and personally identifiable data [27]. Individuals' personal data and information are often time captured and harvested by these corporations without their knowledge and approval. These 'GAFAM' big tech companies are in a unique position to collect, harvest and analyse data generated through the online activity of the individual user on their platforms [34]. They can gain crucial insights into individuals' behaviour and online content consumption. There are also concerns that GAFAM's harvesting of data can open the door to manipulation of online public discourse. This issue gets to the heart of why digital sovereignty is important to an individual. It is an individual's wish, as the data owner, to be able to exercise his/her rights and control the usage and sharing of their data. It is to have the autonomy [32] to decide the destiny of one's own data and content and be able to self-determine whether to participate to share the data for the better good of research outcomes and advancement.

Another fundamental aspect of regaining control of an individual's data is that of identity management for individuals [27]. Identity is key to putting access control policies in place, identifying where data assets are stored, what can be accessed, and for establishing trust between parties. To an individual, the security and privacy protection of their online identity is of particular concern and is an individual's wish to be able to exercise sovereignty. To address the increasing need for online identification, there has been proposed creation of an eID (electronic identity or digital identity) by governments, particularly in EU, which allow citizens to have greater privacy as users of online platforms. However, the current proposal is mostly for access to digital government services [37], while the authentication in the private domain is still being left to the major foreign platforms, such as Facebook, Apple, Amazon, Google, Alibaba, or Tencent. In a centralized identity management system, individuals need to rely on a central authority to keep safe their personal information that is used to verify their online identities. Cases of identity theft are prevalent in the past few years when identity data stored on central authority platforms are stolen and used for malicious purposes. On top of this, the central authority can also monitor the online transactions of the individual based on the identity verification requests that it receives for that individual. This creates large concentrations of personal data and collections of their online activities on these platforms, which has a direct impact on individuals' privacy and digital sovereignty.

In the context of sharing personal and sensitive data for usage by artificial intelligence and machine learning application [28] for the advancement of research and innovations, privacy and data security, and the ability for individuals to control the level of access and disclosure of details are a deterrence to individual's decision to share. To ease the concerns, there are privacy-preserving data mining (PPDM) methodologies [14] to protect and preserve the privacy of data owners. The PPDM methods are designed to guarantee a certain level of privacy, while maximising the utility of the data, such that data mining can still be performed on the transformed data efficiently and yet still preserve the privacy of the data owner. The privacy-preserving methods work by withholding sensitive information about the data owner, and thereby expressing the information selectively for usage. Other than privacy preservation methods, there are also privacy-enhancing techniques that enhance the privacy of personal and sensitive data while also providing useful utilization of these data for machine learning and computation. Some of these privacy-enhancing techniques are Secure Multi-Party Computation (SMPC) [39], Differential Privacy (DP) [15], Fully Homomorphic Encryption (FHE) [16] and Zero-Knowledge Proof (ZKP) [40]. Though these privacy-preserving and enhancing methods and techniques are available to protect data owners' privacy, the level of privacy is usually not decided by them but by data custodians without needing any consent from the data owners [41]. Most of the time, this does not align well with the privacy level that an individual, as a data owner, desires or is comfortable with. The data being shared by individuals can contain identifiable personal data that can relate and link to their personal lives and thus traceable to their online activities. Data owners should have the rights and authority to control the level and details of their own identifiable personal data to be shared. In addition, data owners should have the autonomy to decide whether they are willing to participate in the sharing, the destination of their data and how they should be used or handled. The intention of sovereignty is not to have individual's data records locked away in fragmented organizational silos, not easily accessible and resulting in reduced access and utility, but to leave it to individual, as the rightful owner of the data, to have the sovereignty to control and decide.

In addressing the concerns of the governance of one's digitalized data, a new category of digital sovereignty claim has emerged in recent years. Emphasizing the desire and need of individuals to take control and claim back their sovereignty over the management of their digital identity, personally identifiable information (PII) as well as traits and behavioral data that can identify the individual's online digital self, a category on self-sovereign identity (SSI) was proposed [27]. SSI is a paradigm to decentralize the storage and management of individuals' identity and credentials, and allow individuals to maintain control over their identity across all different services and thus achieve autonomy in managing these services. To have an understanding of what is self-sovereign identity as a form of digital sovereignty and the existing research in this area, a systematic literature review of SSI is conducted to compile an in-depth study and analysis of the SSI paradigm.

## IV. SELF-SOVEREIGNTY IDENTITY

Self-Sovereign Identity (SSI) is a decentralized approach that empowers individuals to have complete control and ownership of their data and be able to decide what data to share and with whom. It enables an individual to have autonomy and control over their identity and personal data with self-determination on the level of data to be shared, used and handled. SSI research has gained much research focus and interest in recent years and is gathering momentum in academic and industry, specifically in computer science. Therefore a detailed and comprehensive systematic literature review on SSI is conducted to provide a good understanding of this topic, the current research focus and trend.

### *A. Research Methodology*

This paper based the research methodology on Kitchenham's guideline [42] using software development lifecycle (SDLC) methodology to provide a systematic literature review (SLR) of SSI by first giving background information on SSI before taking on SDLC phases perspectives. This work thus adopted the SDLC protocol shown Fig. 3 and aligned the researches questions according to the 3 phases: Requirement Analysis, Architectural Design and Implementations.

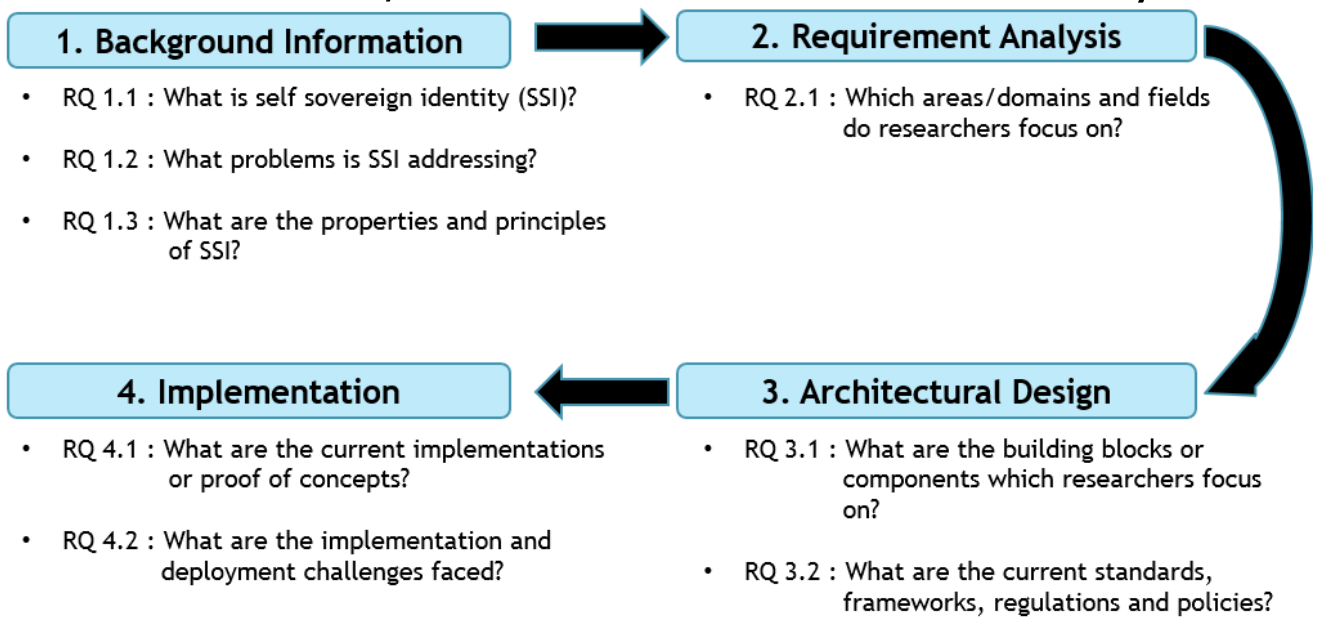

**Fig. 3.** Paper search and selection process map

### B. Research Questions

In view of various information, entities, stakeholders and technologies in a SSI deployment, a background information of SSI is first provided before proceeding to provide a systematic literature review from a software engineering perspective and study SSI from the various SDLC phases; namely requirement analysis, architectural design and implementation. As shown in Fig. 3, this paper adopts the SDLC phases and group the research questions according to each phase. The explanation of how the research questions are derived and elaborated

*a) Background Information*

To understand self-sovereign identity, there is a need to first know the background of self-sovereign identity and the problems it is addressing. Thus, there are RQ 1.1 (What is self-sovereign identity (SSI)?) to understand the definitions of SSI provided by researchers and RQ 1.2 (What problems is SSI addressing?) to know the benefits and objectives of SSI solution. After learning the background and objectives of self-sovereign identity and overview of the problems it is addressing, RQ 1.3 (What are the properties and principles of SSI?) lists and explains the various principles and properties used in this field as elaborated by the gathered papers.

*b) Requirement Analysis*

Knowing the problems SSI addresses, RQ 2.1 (Which areas/domains and fields do researchers focus on?) finds out the areas and domains as well as the involved issues researchers are focusing and working on.

*c) Architectural Design*

With an overview of how SSI works in the requirement analysis, focus is next on the non-functional requirements (since functional requirements are application-specific) to look at the architectural design of SSI building blocks and components to understand what researchers focus on in their design. For this, RQ 3.1 (What are the building blocks and components which researchers focus on?) is derived. As architectural design may need to conform to industry standards and align with defined regulations and policies, and established frameworks may exist for adoption, there is thus, RQ 3.2 (What are the current standards, frameworks, regulations and policies?). These RQs aim to identify the possible approaches that address the non-functional requirements during development and extract the software components for the architecture design to fulfill the non-functional requirements.

*d) Implementation*

In the implementation phase, there is RQ 4.1 (What are the current implementations and proof of concepts?) to know the current existing implementations of SSI. In this phase, there will be implementation challenges faced, thus RQ 4.2 (What are the implementation and deployment challenges faced?) is derived to identify possible challenges to address.

### C. Sources Selection and Strategy

This paper searched through the following search engines and databases: (1) ACM Digital Library, (2) IEEE Xplorer, (3) ScienceDirect, (4) Springer Link and (5) ArXiv. Fig. 4 shows the adopted paper search and selection process.

*a) Search String Definition.*

The following search terms were used to select the initial studies:

("self-sovereign identity" OR "self sovereign identity" OR "digital identity")

The single term "sovereign" and abbreviation 'SSI' were omitted as they can be related to other fields, for example sovereign funds in the financial field and surgical-site infections (SSI) in the medical field. To increase the number of selected studies, the terms "digital identity" which is closely tied to concepts of self-sovereign identity is also included whereby the papers can later be filtered according to the papers' relevance. And since it is a new subject, the period was restricted to studies published between 2017 and 2021. Table I summarizes the captured studies in the searches conducted on Nov 25th, 2021.

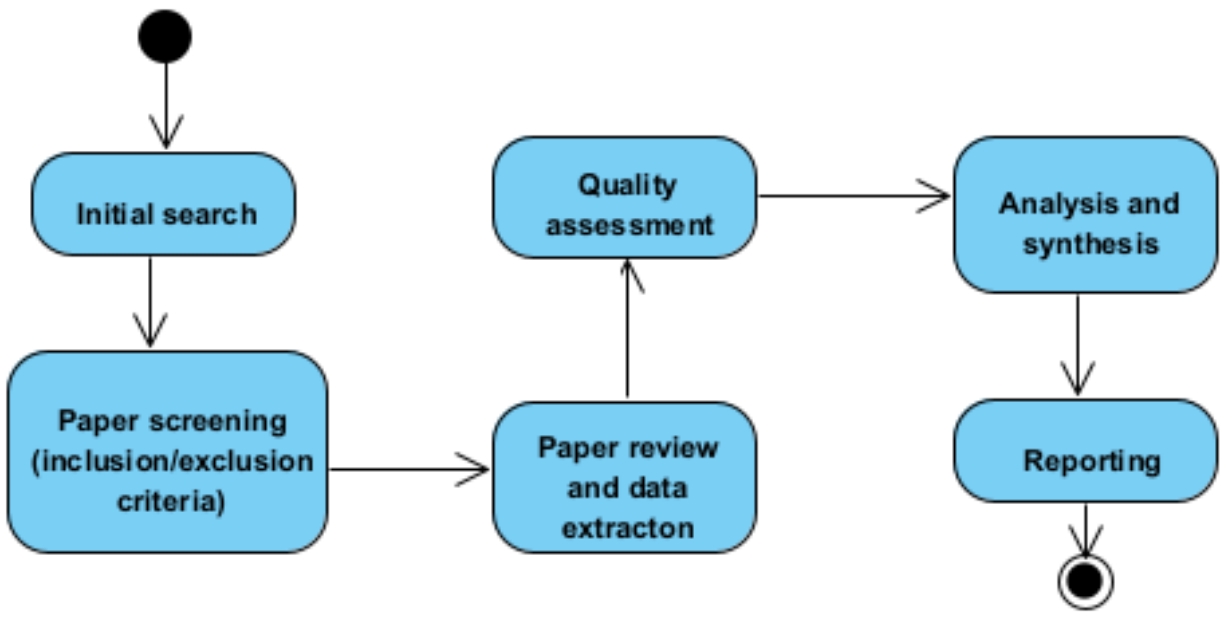

**Fig. 4.** Paper search and selection process map

*b) Inclusion and Exclusion Criteria*

The inclusion and exclusion criteria are formulated to effectively select relevant papers. The inclusion criteria restrict the scope of the selected studies to align with SLR research questions while the exclusion criteria remove unwanted studies in terms of irrelevant types, language and related subjects.

The finalised inclusion criteria are as follows:

- Long and short papers that have elaborated on self-sovereign identity and related systems, specifically focus on the research works that provide comprehensive explanations on its components and functionalities.

• Survey, review, and SLR papers that identify the open problems and future research trends objectively.

The finalised exclusion criteria are as follows:

• Non-primary, short (less than 5 pages) and non-English studies.

• PhD dissertations, tutorials, editorials and magazines.

*c) Quality Assessment*

A quality assessment scheme was developed to evaluate the quality of the papers. There are four quality criteria used to rate the papers:

- The citation rate. This is identified by checking the number of citations received by each paper according to Google scholar.
- The methodology contribution. The methodology contribution of the paper is identified by asking 2 questions: (1) Is this paper highly relevant to the research? (2) Is there a clear methodology that addresses its main research questions and goals?
- The sufficient presentation of the findings. Each paper is evaluated based on the availability of results and the quality of findings (Are there any solid findings/results and clear-cut outcomes?).
- The future work discussions. Each paper is assessed based on the availability of discussions on future work.

TABLE I

NUMBER OF SELECTED PUBLICATION PER SOURCE FOR ON SELF-SOVEREIGNTY IDENTITY

| Sources | ACM | IEEE | Springer | ScienceDirect | ArXiv | Total |
|---|---|---|---|---|---|---|
| Paper count (initial) | 26 | 94 | 74 | 37 | 32 | 263 |
| Paper count (filtered) | 11 | 36 | 31 | 6 | 13 | 97 |

*d) Data Extraction and Synthesis*

All the selected papers are downloaded and the essential information are recorded in a data extraction sheet, including the title, source, year, paper type, venue, authors, affiliation, the number of citations of the paper, the answers for each RQ, and the research classification.

*D. Results*

After executing the source selection and strategy, a total of 97 papers are identified [43-139] for the study. Table I shows the sources of the paper and Fig. 5 shows the paper count by year and the paper type. From the figure showing the paper count year, the growing interest in SSI research can be observed. And most of the papers focus on proof of concept and prototype to elaborate on their proposal to implement SSI solutions.

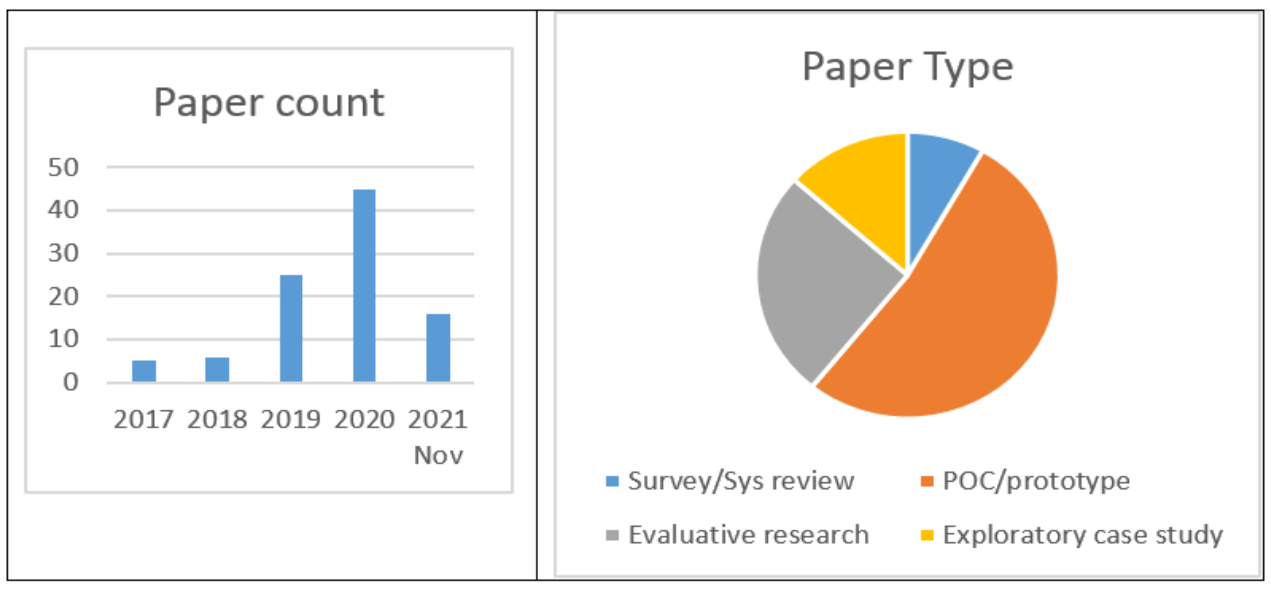

**Fig. 5.** Paper count by year and paper type

*a) RQ 1.1: What is self-sovereign identity (SSI)?*

The first research question (RQ 1) is "What is self-sovereign identity?" To answer RQ 1, the definition and breakdown of the words in self-sovereign identity reported by each study are recorded. This question helps the audience to understand: (1) what self-sovereign identity is, and (2) the perceptions of researchers on self-sovereign identity. To get an overview, a word cloud shown in Fig. 6 is generated and shows the frequency of the words that appear in the studied papers [51, 60, 84, 92, 97, 108-111, 113, 114, 120, 138] which provide an elaborative definition of "self-sovereign identity" in their study. The most frequently appeared and relatable words, other than the self-sovereign identity words, include: individual, control, data, user, digital, principles, consent, credential, ability, disclosure, authority, autonomy, security, single, personal, ownership, portability, transparency, minimal, correlation, ecosystem, etc.

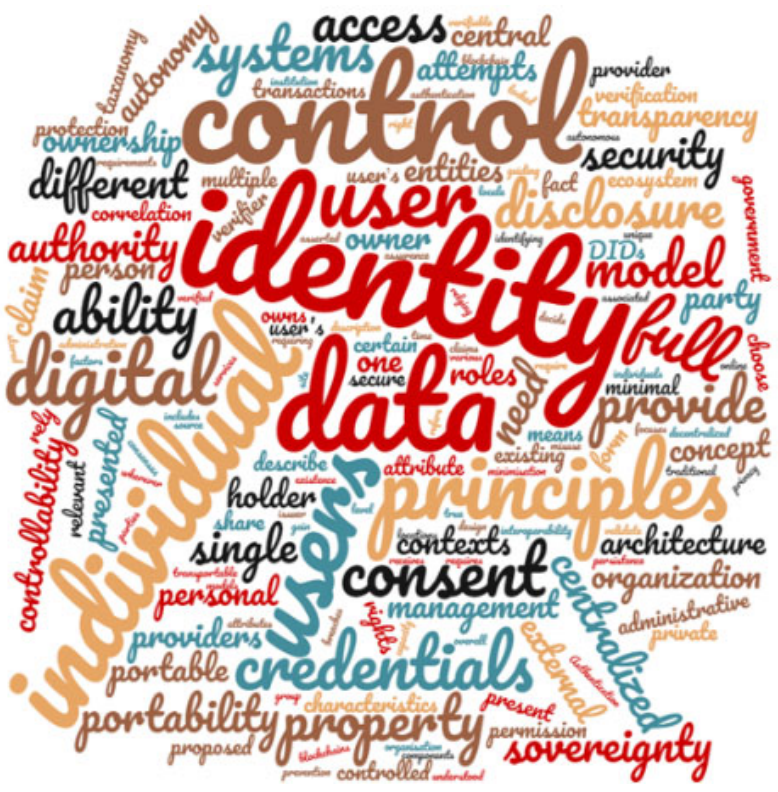

**Fig. 6.** Word cloud for SSI papers

*b) RQ 1.2: What problems is SSI addressing?*

SSI is seen as a solution to move away from a centralized model to a user-centric decentralized model. With the shift, it attempts to address the following problems:

1) Identity theft and fraudulent transaction [48, 53, 57, 62, 64, 72, 73, 75, 77, 80, 81, 85, 87, 89, 90, 96-98, 101, 111-113, 115, 117, 118, 123, 131, 134, 136-138] whereby a central data source captures the personal identifiable information of individual's citizenry, biometric data and private data.

2) Alignment to local regulation and policy on managing data privacy [70, 71, 74, 77, 78, 81, 82, 85, 91, 92, 105, 108, 111, 113, 132, 137].

3) The proliferation of passwords [43, 53, 55-58, 60, 64, 68-71, 74, 75, 77, 78, 81, 88, 90, 96, 98, 101, 104, 106, 108, 109, 111, 113, 116, 133] stored in multiple systems and the risks involved. It is a move towards password-less authentication.

4) Providing digital identity as a form of verifiable document, for example as a national identity, for employment verification and even for refugees in crisis-prone countries [73, 89, 97, 131].

*c) RQ 1.3: What are the properties and principles of SSI?*

The 10 principles of self-sovereign identity written by C.Allen [140] which he outlined in 2016 as the first draft of the concept, was a commonly cited article in the studied papers [49, 50, 52, 53, 55, 60, 61, 67, 71, 75, 77, 79, 80, 83-87, 91, 97, 98, 100, 105, 106, 108-110, 114, 115, 120, 125, 131, 132, 134, 136, 137]. The 10 principles are existence, control, access,

transparency, persistence, portability, interoperability, consent, minimalization and protection. Additional principles of provable, is also proposed by [53, 69, 77, 84, 99]. The objective of each principle is outlined below:

1) Existence. Users must have an independent existence.
2) Control. Users must control their identities.
3) Access. Users must have access to their own data.
4) Transparency. Systems and algorithms must be transparent.
5) Persistence. Identities must be long-lived.
6) Portability. Information and services about identity must be transportable.
7) Interoperability. Identities should be as widely usable as possible.
8) Consent. Users must agree to the use of their identity.
9) Minimalization. Disclosure of claims must be minimized.
10) Protection. The rights of users must be protected.
11) Provable. Claims must be shown to hold true.

From a SSI solution perspective, [110] proposes an extended set of 20 principles to cater to the evolving SSI requirements and standards used to analyse SSI solutions. These 20 principles are relatable to those provided by C. Allen but with the addition of the following principles which relate to the SSI infrastructure and services, other than security, privacy, decentralized, availability and scalability:

1) Recovery. Mechanisms must be in place to recover and re-assert identity due to complete loss of credential.
2) Cost Free. Owning an identity should be free of cost or with negligible cost.
3) Sustainable. An identity infrastructure and services should be environmentally, economically, technically and socially sustainable for the long term.

On categorizing the principles, Sovereign Foundation categorized Allen's principles into three groups; security, controllability, and portability [50], while [60] analyzed existing definitions, extracted properties and classified them into five categories, foundational, security, controllability, flexibility, and sustainability.

*d) RQ 2.1: Which areas/domains and fields do researchers focus on?*

To further breakdown the gathered papers' research focus in term of areas/domains so as to provide an understanding of the current popular research area, the following results on the areas/domains and the papers are gathered:

- Financial Banking [62, 63, 81, 87, 100, 101, 124]
- Education and certification [84, 85, 107]
- Healthcare [46, 82, 95, 127, 130, 134-137, 139]
- National, e-Gov [57, 65, 69, 72, 79, 91, 96, 132, 138]
- Transportation [137]
- Internet of Things (IoT) [55, 66, 70, 88, 103, 105, 106, 117, 126, 128]
- Content Management [100, 112, 120, 121, 125]
- SSI framework and components design [43, 47, 49, 54, 58, 59, 67, 75-77, 80, 83, 86, 93, 97, 98, 102, 104, 109, 110, 112, 114, 122]
- Identity Management (IdM) [44, 45, 48, 50-53, 56, 60, 61, 68, 71, 73, 74, 89, 92, 94, 99, 108, 111, 113, 115, 116, 118, 119, 129, 131, 133]

The focus on SSI framework and components design is on the improvement, extension, analysis and evaluation concerning existing and proposed new framework and its components (eg, user model and authentication, verifiable credential, cryptographic schemes, key management, digital wallet). The focus on content management is on resolving issues of digital rights management, data sharing, exchange and trading. The focus on identity management is on resolving issues of digital identity management, authentication, access control and biometrics. The rest focus on the topic of application and proposal of SSI solutions in the domain of financial banking, education and certification, healthcare, transport, national (e-government) and internet of things. For financial banking, it is for the improvement of the loan application process and verifying asset ownership, a new digital approach to the KYC process and prevention of identity scam, and the use of SSI solution in the rural areas for banking transactions. For education and certification, it is to certify the quality, authenticity of published datasets and sharing the academic credentials. In healthcare, SSI solution is proposed to deal with healthcare data issues for data owners to control and delegate access of personal EHR to relevant stakeholders, eg, a medical practitioner. In transport, a travel credential for taking transportation and eventual improvement to the travel process is proposed. In national (e-government), a verifiable national identity to support identification, authentication and universal interoperability is proposed. In IoT, a decentralised, transparent digital identity for remote devices for purposes of authentication, verification, and authorization is proposed. Fig. 7 shows the paper focus by percentage.

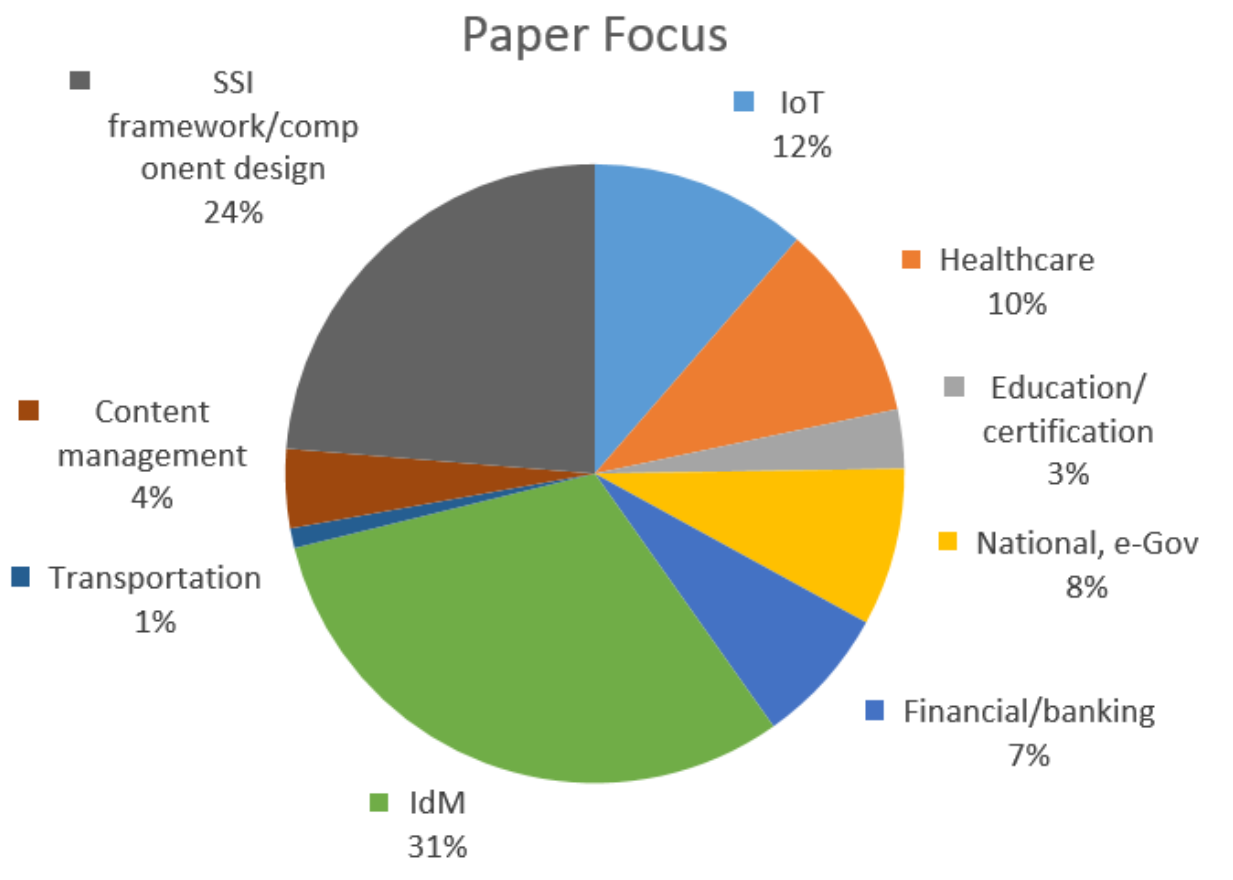

**Fig. 7.** Area of focus of the gathered papers

*e) RQ 3.1: What are the building blocks and components which researchers focus on?*

There are several main building blocks and components in the SSI solution architecture, namely; Verifiable Credential/Claim (VC), Decentralised Identifier (DID), Decentralized Verifiable Data Registries (DVDR), Privacy-Promoting Credential and Claim Checks (PP), Personal Data Stores (PDS), DID Communication (DIDComm), Governance

Frameworks (GF) and decentralized Distributed Ledger Technologies (DLT) or Blockchain. Blockchain is the underlying foundation platform used by most of the gathered papers. Other than papers that focus on the evaluation of various blockchain platforms or Distributed Ledger Technology (DLT), there are also proposed solutions to improve the SSI building blocks and components. Fig. 8 shows the papers’ coverage.

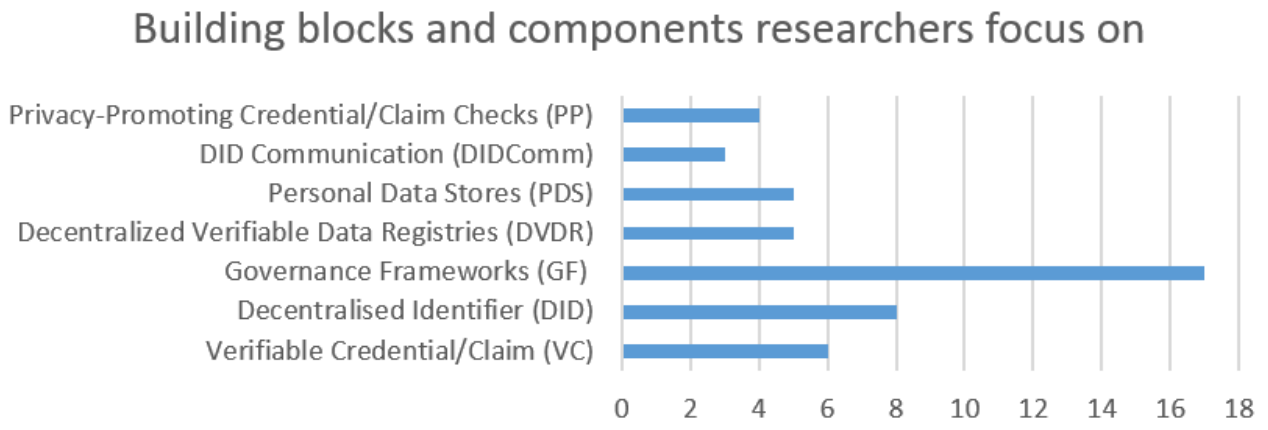

**Fig. 8.** Building blocks and components researchers focus

For VC, [64, 75, 119, 112, 125, 132] look into the backup and restore/recovery of user’s verifiable credential, identity data and cryptographic keys with support of flexible proof mechanisms to ensure the credentials are cryptographically reliable as established by the issuer. For DID, [49, 51, 58, 67, 76, 88, 105, 116] focus on attribute certificate and cryptographic key creation and management and DID search to provide a means for both issuer and verifier to establish the identity of the holder without reliance on a centralized party. In looking at ways to improve governance framework (GF) and its protocols, [44, 45, 54, 56, 62, 69, 81, 84, 90, 93, 101, 104, 106, 107, 111, 117, 126]] look into the authentication, authorisation protocol, identity/credential proofing and verification mechanism, with creation of structures, roles, and policies, adaptable for SSI approaches to different domains. For DVDR, [71, 80, 85, 122, 129] explore the attribute access control and the suspension or revocation of verifiable credentials. On PDS, [52, 94, 118, 121, 137] focus on the VC and DID store on the digital wallet and passports to support individual control over sharing and access to these data. On DIDComm, [78, 128, 131] focus on the communication channel to facilitate DID sharing and verification with regard to the acquisition, processing, and distribution of personal information. On PP, [63, 99, 122, 123] look into privacy preservation techniques, specifically zero-knowledge proof schemes, inspect the current status of a credential without revealing any additional personally co-relatable data about the individual.

*f) RQ 3.2: What are the current standards, frameworks, regulations and policies?*

There are various standards, frameworks, regulations and policies in support of the SSI architecture and solution as well as conformance to the government legal framework. A consolidation of the SSI-related standards, frameworks, regulations and policies, as were mentioned and discussed in the studied papers, are provided in tabulated form in TABLE II, III and IV respectively.

*g) RQ 4.1: What are the current implementations or proof of concepts?*

In moving towards a password-less authentication, [116] implemented and evaluated a password-authenticated decentralized identity (PDID) framework providing password-authenticated decentralized identities with global and human-meaningful names. It uses the OPAQUE protocol, a secure asymmetric password authenticated key exchange (aPAKE) and provides the performance metric of PDID operations. [45] explored the capability of current smartphones to acquire, process and match fingerprints using only its built-in hardware and devised a mobile biometric-based authentication system that only relies on local processing without requiring any cloud service, server, or permissioned access to fingerprint reader hardware. It proposed it as a key building block for a self-sovereign identity solution that integrates permission-less blockchain for identity and key attestation.

For secure decentralized identifier communications and exchange of attributes, [126] presented a solution for self-sovereign identification and communication of IoT agents in under constrained networks using SWARM framework. It implemented an optimized layer for the protection of messages exchanged between self-sovereign agents which uses a binary encoding that can achieve size reduction for signed and encrypted messages. [128] developed a hardware solution based on ESP32 using SX1276/SX1278 LoRa chips, with adaptations made to the lmic- and MbedTLSbased software stack to solve the challenge of verifying the identity of nodes within a wireless ad-hoc mesh network and the authenticity of their messages in sufficiently secure, yet power-efficient ways. To address the limited ATM facilities in rural areas, the high initial cost of ATM deployment, potential security issues in ATM systems and high inter-bank transaction fees, [124] implemented a blockchain-based, low cost, peer-to-peer money transfer system as an alternative for traditional ATM system and debit/credit card system. It also provided a self-sovereign identity empowered mobile wallet for its end users.

TABLE II
STANDARDS REFERENCED BY GATHERED PAPERS

| Standards | Description | Origin |
|---|---|---|
| Biometrics Open Protocol Standard (BOPS) | A standard for a software-based system for identity assertion. | IEEE |
| Fast Identity Online (FIDO) | An open authentication standards the enable a service provider to leverage existing technologies for passwordless authentication. | Alliance |
| Federated Identity Management (FIdM) | A Federated SSO to establish a trusted relationship between separate organizations and third parties to share identities and authenticate users across domains. | |
| Identity-Mixer | A cryptographic protocol suite for privacy-preserving authentication and transfer of certified attributes. | IBM |
| Open Authentication (OAuth) | An open standard for access delegation used as a way for Internet users to grant websites or applications access to their information on other websites without giving them the passwords | IETF |
| OpenID Connect (OIDC) | An open authentication protocol that profiles and extends OAuth 2.0 to add an identity layer. | Foundation |
| Security Assertion Markup Language (SAML) | An open standard that allows identity providers (IdP) to pass authorization credentials to service providers (SP) | OASIS |
| U-Prove | A SDK for user-centric identity management which enables application developers to reconcile security and privacy objectives (including anonymity), and allows for digital identity claims to be efficiently tied to the use of tamper-resistant devices | Microsoft |
| W3C Verifiable Credentials | Specifications for an open standard for digital credentials to make expressing and exchanging credentials that have been verified by a third party easier and more secure on the Web. | W3C |

TABLE III
SSI RELATED FRAMEWORKS

| Frameworks (SSI) | Description | Origin |
|---|---|---|
| European Self Sovereign identity framework (ESSIF) | part of the European blockchain service infrastructure (EBSI) to provide an interoperable and decentralized framework for people to control their own identities across EU member states, without having to rely on each individual government | EU |
| **Frameworks (Blockchain)** | **Description** | **Origin** |
| Blockcert | an open standard for creating, issuing, viewing, and verifying blockchain-based digital records registered on the blockchain | |
| Civic | a blockchain-based identity management solution that gives individuals and businesses the tools they need to control and protect personal identity information | Commercial |
| Hyperledger Indy, Aries, Ursa | Hyperledger Sovereign Identity Blockchain Solutions. Indy is a distributed ledger, purpose-built for decentralized identity. Aries provides an infrastructure for peer-to-peer network and blockchain-rooted interactions within platforms to offer key management and secret management systems. Ursa provides cryptographic support | Open source |
| Jolocom | open source protocol for decentralized digital identity and access right management. | Open source |
| Shocard (PingID) | stores one’s identity onto bitcoin’s blockchain so that one can prove his/her identity whenever needed. | Commercial |
| Sovrin | an identity metasystem for SSI with a public service utility enabling self-sovereign identity on the internet. | nonprofit organization |
| uPort | a self-sovereign identity and user-centric data platform on Ethereum | donated to the Decentralized Identity Foundation |
| Veres One | a blockchain for acquiring and managing decentralized identifiers. | Global network for identity |

TABLE IV
REGULATIONS AND POLICIES

| Regulation and policy | Description | Origin |
|---|---|---|
| California Consumer Privacy Act (CCPA) | To protect the data privacy rights of citizens living in California. | US |
| Convention on the Rights of the Child 41(CRC, non-US) | The most universally accepted human rights instrument, ratified by every country in the world except two | UN |
| electronic identification and trust services (eIDAS, EU) | An EU regulation on electronic identification and trust services for electronic transactions in the European Single Market | EU |
| European Convention on Human Rights (ECHR), Article 8 | To protects the human rights of people in countries that belong to the Council of Europe | EU |
| General Data Protection Regulation (GDPR, EU) | A regulation in EU law on data protection and privacy in the European Union (EU) and the European Economic Area (EEA) | EU |
| Health Insurance Portability and Accountability Act of 1996 (HIPAA) | A federal law that required the creation of national standards to protect sensitive patient health information from being disclosed without the patient’s consent or knowledge. | US |
| International Covenant on Civil and Political Rights (ICCPR, US) | Commits its parties to respect the civil and political rights of individuals, including the right to life, freedom of religion, freedom of speech, freedom of assembly, electoral rights and rights to due process and a fair trial | UN |
| Payment Services Directive Two (PSD2) | To make online payments safer for customers, improve protection of consumer information, address payment fraud, and provide a common platform for competitors | EU |

In the management of DID and user attributes, [51] implemented a decentralised service for self-sovereign identity management which has user-managed attributes using a decentralised query protocol without a central party, and with privacy-preserving features such as ABE-based access control to sensitive attribute data. To address the issues of data ownership and isolation, lack of accountability, and high privacy risks existing in current EHR systems, and allow patients to have control over their personal health data, [72] implemented a mobile healthcare system for personal health data collection, sharing and collaboration between individuals and healthcare providers, as well as insurance companies. Its implementation adopted blockchain and Intel SGX technology. [58] implemented a decentralised platform that allows for issuing, managing and verifying digital documents using the Ethereum blockchain which also supports attribute-based authentication. It facilitated the verification of the origin and integrity of paper documents as well as the recovery of lost documents. [122] implemented a blockchain based self-sovereign identity management (BbSSIM) technology and used threshold CP-ABE to achieve access control to resolve issues of user attribute revocation existing in multi- attribute authority (AA) and threshold multi-AA schemes deployed in large-scale cloud or cross-cloud access. To address the need for trusted third party KGC (Key Generator Center) to generate and manage the keys for the user, [76] proposed a decentralized identity authentication and key management scheme using blockchain technology.

In addressing privacy and security requirements for users’ digital identity information without a centralized party managing the identity exchange transactions, [63] presented a decentralized protocol for privacy-preserving exchange of users’ identity information and digital assets using a permissioned blockchain network. To address the major concern over health data privacy of the data collected from wearable devices, which can reflect patients’ health conditions and habits, and the increased data disclosure risks among the healthcare providers and application vendors, [46] implemented a decentralized blockchain and made use of the trusted execution platform enabled by Intel SGX to provide accountability for data access. [79] implemented a decentralized eID derivation system that enables users to selectively disclose only relevant parts of the imported identity assertion according to the services’ requirements. This addresses the lack of qualified identity data that satisfies the services’ requirements and also protection of users’ privacy during the derivation of electronic identity (eID) data. [123] leverages the smart contracts and zero-knowledge proof (ZKP) algorithms to implement a system prototype named BZDIMS that includes a challenge-response protocol and allows users to selectively disclose their ownership of attributes to service providers to protect users' behavior privacy. To address the privacy concerns of sharing digital credentials for secure verification of participants' identities and credentials to increase trust, and also allow data minimisation mechanism to reduce the risk of oversharing the credential data, [107] implemented a blockchain-based Self-Sovereign Identity (SSI) platform architecture that allows secure creation, sharing and verification of credentials. [121] implemented a self-sovereign blockchain-based privacy-preserving matchmaking platform that enables its users to treat their personal data as a digital asset and trade it according to the matching score with other users. In the light of the COVID-19 pandemic and the need for automated and efficient human contact tracing that would be non-intrusive and effective as well as in a privacy-preserving fashion, [82] developed "Connect", a Blockchain and Self-Sovereign Identity (SSI) based digital contact tracing platform. [85] developed a Self-Sovereign Identity Based Access Control model for cross-organization identity management to show how SSI can fit within the context of established enterprise identity and access management technologies to alleviate data breach and user privacy problems.

In the area of personal data store, to address the closed, proprietary data silos of existing web-based social communication platforms which not only lock in users into their service platforms, but also control and exchange personal and sensitive data such as photos, messages, or contact information, [78] implemented a decentralized service architecture, using Ethereum blockchain, for users to take full control of their personal data, which is stored and managed by personal data storages. And to address the backup and restore issue of user's identity data, [64] combines SSI sustaining aspects and extends them to create a backup-and-restore protocol with authenticated backup and auditing by remote entities.

*h) RQ 4.2: What are the implementation and deployment challenges faced?*

There are several challenges identified and discussed in the studied papers on the SSI implementation and deployment in realizing the SSI solutions and models. The challenges faced by the researchers can be grouped into the following areas:

**Key management**

Decentralized Identifiers (DIDs) face challenges like distribution of public keys, keeping changes made to the keys, storage capacity and scalability [105]. DIDs and the accompanying DID Documents (DDoc) enable individuals to share abstract identifiers (DIDs) with an associated key pair and a resolution end-point. Depending on the individual's need to have DID to verify one's identity credential or claim, each individual can have multiple DIDs with corresponding DDocs and associated key pairs. The identity could use one key pair for authentication, another one for encrypted messages, and another one for its verifiable credentials (VC). The keys can also be defined to permit which keys are allowed to authenticate changes to the DID Document itself. VC issuers and holders may also revoke the VCs if it is no longer valid and this will involve the revocation of the associated keys. Thus maintaining and managing these multiple key pairs is a challenge with regards to the distribution, update, storage, recovery and revocation of the keys that are tied closely to the DID, DDoc and VC [122]. From an IoT deployment perspective, the availability of efficient and less resource intensive open software libraries to facilitate the key management and operation that manipulates DIDs and VCs on the IoT devices [76, 88] is a challenge.

**Security of personal datastore/wallet**

DIDs, DDoc and VCs are stored on individual's personal data store or wallet. Thus the implementation needs to ensure the security of the personal data store and the ability for users to be able to securely manage their own identities. The solution will require to consider an efficient and secure backup, restore and recovery of the user's identity data and cryptographic public keys [49, 64, 75] as well.

**Scalability and reliability of solutions**

Distributed ledger technology (DLT) is used as an immutable transparent shared storage whereby transaction records and data are accessible by parties in the ecosystem. Scalability and reliability are intrinsic challenges of DLTs themselves [48]. Blockchain is the most common type of DLT used in the studied papers for the implementation and deployment of SSI solutions. Other than in the area of cryptocurrencies, blockchain has yet had a wide real-world practical application deployment due to these mentioned challenges.

**Trust and assurance**

For wide deployment of SSI solutions for identity management with eID, the build-up of a chain-of-trust connecting various eID systems via the SSI ecosystem with different stakeholders [79] is necessary. This requires buy-in and adoption from stakeholders to participate and build up the trust to evolve the ecosystem [93]. From users' perspectives, data link to their identity and PII may be stored on DLT and private information may be shared, thus, in order for users to participate in the SSI ecosystem, there must be means for users as individuals to verify that their privacy are protected in terms of ensuring confidentiality, anonymity and unlinkability [63] of their identities.

**Standardization**

The fragmented nature of SSI market, the immaturity and incompatibility of standards, legal and regulatory uncertainty have an impact on SSI ecosystems and their possible adoption [131]. In terms of identity management (IdM), the design of IdM must be based on open standards and established protocols to ensure maximum transparency and adoption [92]. The assurance of interoperability of the solution in real world applications and backward compatibility are also much needed [113] to ensure the sustainability of the solution.

*E. Analysis and discussion.*

The research on Self-Sovereign Identity requires a good understanding of these three words and addressing the needs to fulfill the objectives and definition of these words. Gathering the definition provided by the studied papers, the individual word can be defined as:

- **Self**: an individual, own, personal, user-centric;
- **Sovereign**: independent, self-governing, control, authority, autonomy, empowerment, without asking for permission;
- **Identity**: a distinguishing character, personality, attributes and behaviors by which a person or thing is recognized.

Self-sovereign identity can be viewed from both a conceptual and technical solution perspectives.

The conceptual motivation of self-sovereign identity is to empower an individual with the autonomy and authority to control his/her own personal and distinguishable data. The empowerment allows individuals to take ownership of his/her data and independently grant his/her authorization and consent on the disclosure of his/her personal data based on their principles - personal control to protect one's privacy.

From a technical point of view, a self-sovereign identity solution is to build a decentralized user-centric ecosystem that grant an individual user or entity with the rights to control, authorize and consent on the disclosure and usage of its own digital identity or credentials in order to fulfill a transaction. The digital identity which is a user's personally identifiable information (aka a user's digital twin) can be in the static data form like credential (example, user identity, password and social security number) and personal attributes (trait like height, weight, hair color, biometric characteristic) or dynamic data like individual behavior (in the form of gait like walking posture, strength of keystrokes, speaking tones).

To an individual, data security and privacy are of utmost importance. Thus most of the studied papers focus on addressing problems on the protection of an individual identity against theft and exploitation. The other common problem is the proliferation of one's passwords against multiple online platforms that are a common target for malicious entities. SSI concepts of decentralizing the storage of these credentials in the form of verifiable credentials, issued by authorized and trust issuers, stored within one's digital wallet enables the individual entities to control the usage of their own data in a decentralized ecosystem. There are varying properties and principles defined for SSI that attempt to serve as a bounding guide for proposed SSI solutions in the execution of self-sovereignty and protection of an individual's identity. Controllability and security are prominent as the principles proposed by researchers as well as sustainability. In terms of wide adoption, portability and interoperability are also desired properties which currently are also challenges faced in order to have a wider deployment.

Though Wide Web Consortium (W3C) [141] and Decentralized Identity Foundation (DIF) [142] strive to establish standardization for a new decentralized identity ecosystem, however, SSI is still in its infancy of establishing standards with still ambiguity, misunderstanding, and disagreement about key concepts' definitions along with workable and adaptable frameworks. The standards are created to establish interoperability and portable of the identity and credentials and to facilitate the exchange and management of identities and credentials. The deployment of SSI solutions and architectures requires the use of distributed ledger technology (DLT) as an essential building block, and blockchain is the most preferred DLT platform for the development and deployment of SSI solutions. And to facilitate the creation of SSI ecosystem, frameworks are built to support the development of SSI solutions. Among the frameworks, the most commonly used SSI framework is HyperLedger Indy (with the close support of its sibling frameworks; Aries and Ursa), likely due to the blockchain foundation established by the Linux Foundation and its members as well as the HyperLedger Fabric community. Sorvin and Uport are also common among the studied papers as a SSI blockchain platforms for the deployment of decentralized identities and verifiable credentials.

SSI is still an emerging trend, much attention of researchers proposed solutions and works are on the design of governance frameworks to establish a secure, reliable protocol and mechanism by creating structures, roles, and policies for organizations or governments to facilitate the adoption and adaption of SSI approaches to different domains which can resolve questions of trust within different stakeholder groups. DID is also a focus as it is associated to document which holds critical contents to establish the authenticity and verifiability of claims and identity, while at the same time protecting the privacy of the holder through selective disclosure on a need-to-know basis. This is of utmost concern to an individual in protecting his/her identifiable information but at the same time able to provide the needed authentication and verification. The suspension, revocation and recovery of VC and DID, both on individual's personal storage or issuer's data registries are also of major focus for researchers.

Though current SSI researches have presented proposals and solutions to resolve the various challenges and gaps in the current SSI developments and deployments, but through the elaboration of RQ 4.2, there are still open problems. From computer science perspectives, open problems in regards to the portability and interoperability of the solutions, miminalization, protection and provability of verifiable claims, and a better distributed key management system (DKMS) remains to be solved. And from social and national perspectives, legal supports (including regulation and policy), as well as institutional, organizational and user adoptions are key challenges in SSI deployment.

## V. Related Work

The objectives of this systematic literature review paper are to study and provide insights into the research papers that have been written on the topic of sovereignty and its coverage, particularly focusing on self-sovereignty identity. Survey papers are included in the study to provide a good sense of the interests in this area. Thus, for related work, three papers that also provide either a systematic literature review or mapping are identified. [136] provided a systematic literature review focusing on healthcare to investigate state-of-the-art measures based on SSI and blockchain technologies for dealing with electronic health records (EHRs), identifying gaps, and determining the key questions for future research. It looked at questions in the area of the access control mechanisms for EHR access and sharing, the privacy and security risks regarding unauthorized medical information disclosure and its storage. It concluded that it is still a novel subject that medical institutions have been exploring for some time, but still face the challenge of breaking the silos of health information and privacy concerns. Blockchain technologies have provided a viable means of addressing the fundamental challenges of EHR solutions such as access control, data integrity, interoperability, and auditing. The principles of self-sovereign identities (SSI) could be adopted to provide patient-centric healthcare solutions to ensure that patients, as the data owner, have complete control over their data. However, the paper's coverage is mainly on healthcare and SSI adoption in this domain and does not cover other domains like this paper. [143] presented a systematic mapping and systematic literature review covering theoretical and practical advances in Self-Sovereign Identity. As part of their research questions on SSI, it identified the authors of the papers, the co-references and co-authorship in this SSI ecosystem, the conceptual ideas introduced and refuted, a formal definition of SSI and the practical problems that have been introduced and solved. It also attempted to introduce mathematical formulations to precisely define one or more SSI-related problems and presented a solution to the pragmatical problem related to the SSI ecosystem. [144] presented a systematic mapping methodology to provide a coarse-grained overview of decentralized and Self-Sovereign Identity and structure the research area by identifying, analyzing, and classifying the research papers according to their contribution, application domain, IT field, research type, research method, and place of publication. In contrast to these two review papers, this paper provides a wider study to include data and digital sovereignty to lay a background for SSI and also to establish an overview understanding of sovereignty as a whole on digitalized data.

## VI. CONCLUSION

This paper conducted a systematic study of data and digital sovereignty as well as self-sovereign identity and presented the findings, observations with analysis and discussions.

Data sovereignty is not a new research topic as research had been performed by humanity and sociological researchers to look into the concerns and wishes of the indigenous peoples to protect their land, cultural heritage and assets. The CARE and FAIR principles are respectively defined to advance the legal principles underlying collective and individual data rights in the context of the UNDRIP [23] as well as to support data and knowledge integration and promote sharing and re-use of data. With the movement of digitization to turn analogue data to digital, nation-states start to be aware of the need to have sovereignty over their citizens' data to protect their well-being and privacy. Regulations, like the European Union's General Data Protection Regulation (GDPR), policies and laws are created to protect the data, specifically how data are used, where they are stored and their physical location. At its core, data sovereignty is about protecting sensitive, private data and ensuring it remains under the control of its owner within the specified country. However, with the lack of operational details on implementation mechanisms and enacting systems to execute and realize the data sovereignty principles and law, it is a challenge to govern the stewardship and application of data to fully assert the sovereignty of the locally hosted data of the people they are designed to protect.

Digital sovereignty is a more recent topic emerging from the acceleration of digital transformation with the advancement in technologies. With the dominant position of big tech companies in the field of cloud computing and social media, data of citizens and companies are virtually stored and utilised in the cloud of these big tech companies. To wrestle back the control and reassert their authority, nations like the EU initiated the GAIA-X cloud project and the European Cloud Federation initiative to create its own European offering of cloud infrastructure where the customer has full control over the storage, access and processing of the data. However, the project requires a high degree of commitment and coordination from the EU member states which can be a challenge. In addition, with the lack of binding European cloud policies, currently the cloud choices made by the EU member states are through outsourcing, which brings about the risk of vendor lock-in that threatens digital sovereignty. On a personal basis, digital sovereignty is also the wish of an individual, as individual users of digital technologies and services, to have the autonomy to determine who gets to access and use the data about them as individuals. To an individual, the protection of their online identity is of particular concern and an individual's wish to be able to exercise sovereignty over one's identity. Cases of identity theft and the ability of central authority platforms to monitor the individual's online transactions have a direct impact on an individual's privacy and digital sovereignty.

With a shift in trend in digital sovereignty in favour of indivduals, the research topic on self-sovereign identity has gained popularity and interest among computer science academia. The concept of self-sovereign identity is to empower an individual to take ownership and control of his/her own personal and distinguishable data and be able to independently authorise the disclosure of his/her personal data so as to protect one's privacy. From a technological angle, a self-sovereign identity solution is to build a user-centric, decentralised ecosystem that empowered an individual user with the ability to control its own digital identity to authorize and consent on its disclosure and usage to fulfill a transaction. Data security and privacy is of utmost importance and focus. There are varying properties and principles defined for SSI to serve as a bounding guide for proposed SSI solutions. For wide adoption, World Wide Web Consortium (W3C) [141] and Decentralized Identity Foundation (DIF) [142] have defined standards for a new decentralized identity ecosystem. SSI is still in its infancy of establishing standards and frameworks, thus research focuses are still on improving the SSI frameworks and components design and the SSI core – identity management and governance. Though current SSI researches have presented proposals and solutions to resolve the various challenges and gaps in the current SSI developments and deployments, there are open problems in regards to the portability and interoperability of the solutions, minimalization, protection and provability of verifiable claims and a better distributed key management system (DKMS). And from social and national perspectives, legal supports (including regulation and policy), as well as institutional, organizational and user adoptions are key challenges in SSI deployment.